\documentclass{article}

\usepackage{graphicx}
\usepackage{wrapfig}

\usepackage{array}
\usepackage{multirow}
\usepackage{tabularx}
\usepackage[para,online,flushleft]{threeparttable}
\usepackage{makecell}

\usepackage{booktabs}
\usepackage{pifont}
\usepackage{lipsum}
\usepackage[numbers]{natbib}
\usepackage{booktabs}
\usepackage{siunitx}

\usepackage{url}

\usepackage{dblfloatfix}
\usepackage{blindtext}
\usepackage{flushend}

\def\BibTeX{{\rm B\kern-.05em{\sc i\kern-.025em b}\kern-.08emT\kern-.1667em\lower.7ex\hbox{E}\kern-.125emX}}

\begin{document}

\title{Advanced Metering Infrastructures: Security Risks and Mitigation}

\author{Gueltoum Bendiab, Konstantinos-Panagiotis Grammatikakis,\\ Ioannis Koufos, Nicholas Kolokotronis, and Stavros Shiaeles%
\thanks{G. Bendiab and S. Shiaeles are with the Cyber Security Research Group, University of Portsmouth, Portsmouth, UK (e-mail \{gueltoum.bendiab,\,stavros.shiaeles\}@port.ac.uk)%
\newline\indent%
K.-P. Grammatikakis, I. Koufos, and N. Kolokotronis are with the Department of Informatics and Telecommunications, University of the Peloponnese, Tripolis, Greece (e-mail \{kpgram,\,ikoufos,\,nkolok\}@uop.gr)%
}}

\date{}

This paper is a preprint; it has been accepted for publication in the 15th International Conference on Availability, Reliability and Security (ARES 2020), August 25--28, 2020, Virtual Event, Ireland (DOI: 10.1145/3407023.3409312).
\bigskip

\noindent{\bf ACM copyright notice}

\noindent%
Permission to make digital or hard copies of all or part of this work for personal or classroom use is granted without fee provided that copies are not made or distributed for profit or commercial advantage and that copies bear this notice and the full citation on the first page. Copyrights for components of this work owned by others than the author(s) must be honored. Abstracting with credit is permitted. To copy otherwise, or republish, to post on servers or to redistribute to lists, requires prior specific permission and/or a fee. Request permissions from \url{permissions@acm.org}.

\maketitle

% Section % ================================================= |
\begin{abstract}
\noindent%
Energy providers are moving to the smart meter era, encouraging consumers to install, free of charge, these devices in their homes, automating consumption readings submission and making consumers life easier. However, the increased deployment of such smart devices brings a lot of security and privacy risks. In order to overcome such risks, Intrusion Detection Systems are presented as pertinent tools that can provide network-level protection for smart devices deployed in home environments. In this context, this paper is exploring the problems of Advanced Metering Infrastructures (AMI) and proposing a novel Machine Learning (ML) Intrusion Prevention System (IPS) to get optimal decisions based on a variety of factors and graphical security models able to tackle zero-day attacks.
\bigskip

\noindent{\bf Keywords:} Cyber-security, power grid, intrusion detection, malware detection, attack mitigation, graphical security models.
\end{abstract}

% Section % ================================================= |
\section{Introduction}
\label{sec:intro}
The so-called ‘smart devices’,an indispensable part of the Internet of Things (IoT), have provided an ubiquity of connected systems aiming at improving the quality of our life \cite{ss1}. An increased number of businesses, homes and public areas are now starting to use these intelligent devices. The number of interconnected IoT devices (wide-area and short-range IoT connections) in use worldwide has already exceeded 10.7 billion since 2019, and is expected to grow to 24.6 billion by 2025 \cite{ss3}.

All security reports warn that more than 80\% of connected smart home devices are vulnerable to a wide range of attacks \cite{ss11,ss25}.  As compared to traditional computers and mobile phones, resource-constrained IoT devices lack computing power, memory and storage \cite{Kouicem}. Some devices could also be deployed in remote locations that depend on battery power. As a result, these resource-constrained devices are unable to process antivirus software and cryptographic algorithms required for essential security protocols, thus increasing their vulnerabilities. Likewise, the interconnected nature of IoT devices could also allow cybercriminals to carry out parallel attacks once they infiltrate a network. For instance, Mirai botnet that appeared in late 2016, is a high profile example of such risk where embedded and IoT devices were used to execute massive distributed denial-of-service (DDoS) attacks on popular services, like Twitter, Netflix and PayPal \cite{Antonakakis}. Similarly, adversaries have also targeted the IoT ecosystem using gateway attacks, side-channel attacks, malicious injection attacks, Sybil attacks, routing attacks and physical tampering \cite{Mosenia}. Lack of robust government policies and agreed upon standards among vendors and standardisation bodies have also resulted in the production of IoT devices with weak security protocols \cite{Hassan}. Instead, vendors are more focused on the rapid and mass production of devices with less concern for security. Likewise, vendors are also not providing enough support to legacy IoT devices with firmware updates and security patches \cite{Ray}, further weakening the defence of IoT ecosystem. As a result, security breaches could also expose the confidentiality and privacy of data, possibly violating legal obligations such as the General Data Protection Regulation (GDPR) and resulting in severe penalties. However, designers, manufactures and policymakers are not the only ones responsible for breach of IoT security. Often, users who lack IT security knowledge use default usernames and passwords, making them easier to hack once identified using tools such as Shodan and IoTSeeker \cite{Kotak}. Although these vulnerabilities highlight the risks and challenges for IoT, security professionals are constantly exploring new ideas to strengthen its cyber-security.

From the early stage, researchers have been investigating numerous novel cyber-security measures focusing specifically for IoT domain as several studies \cite{Mosenia,ss1,Niraja,Khan} indicate that traditional cyber-security measures are not suitable for current IoT networks and devices. Reasons vary from heterogeneous nature of IoT devices that lack computational power \cite{Kotak} to well-known Intrusion Detection System (IDS) such as Snort and Suricata \cite{Kolokotronis1} not optimised for IoT. Hence, researchers, vendors and standardisation bodies are proposing new solutions to tackle the issue with their own unique approaches. For example \cite{Moosavi} are proposing an end-to-end security scheme for IoT-based healthcare systems, using certificate-based Datagram Transport Layer Security (DTLS) and smart gateways. Similarly, \cite{Mukherjee} is suggesting similar end-to-end security using a novel security middleware that keeps track of (D)TLS sessions, Pre-Shared Keys (PSKs), and device IDs. On the other hand, some solutions focus solely on protocols. For instance, wireless protocols such as Bluetooth Low Energy (BLE), ZigBee and IPv6 over Low-power Wireless Personal Area Network (6LoWPAN); and application protocol like Constrained Application Protocol (CoAP) \cite{Al-Sarawi}. Likewise, researchers are suggesting further solutions based on trust management \cite{Kolokotronis2}, access management \cite{Vanickis} and intrusion detection \cite{Chaabouni}. In addition to academics, government bodies and standardisation organisations are also working on developing necessary frameworks and guidelines in order to provide clear directions to manufactures and promote consumer confidence. For example, online Trust Alliance, an initiative within the Internet Society (ISOC) put forward an IoT Trust Framework \cite{Framework} with strategic principles that highlighted support for IoT devices throughout its entire life-cycle. Likewise, European Telecommunication Standards Institute published ETSI TS 103 645 \cite{CYBER}, a technical publication aiming to guide developers and manufactures on ensuring the security of their IoT devices. Although the importance of these frameworks and guidance cannot be denied, it can be argued that manufactures are only likely to comply if policies are made mandatory as such changes can likely increase production cost and impact on their profit margins. Nevertheless, the continuous effort from all stakeholders has certainly made a positive contribution to the security of IoT and researches are continually searching for newer technologies to further improve this process.

This paper aims at addressing the challenges in AMI security by proposing a new Intrusion Prevention System able to detect and mitigate attacks using Machine Learning and Graph Theory for optimal decision on the threat detected.

The article is organized as follows: in section II a high-level overview of AMI is presented; in section III the Cyber Security risks along with their impact at AMI are discussed; in section IV the proposed Machine Learning (ML) IDS, able to detect unknown (zero-day) threats is explained; in section V the intelligent intrusion response system accompanying the ML IDS for optimal decision support is analysed; finally conclusions and future work are reported in section VI.

% Section % ================================================= |
\section{Advanced metering infrastructure}
\label{sec:ami}

\begin{figure}
\centering
\includegraphics[width=0.75\linewidth]{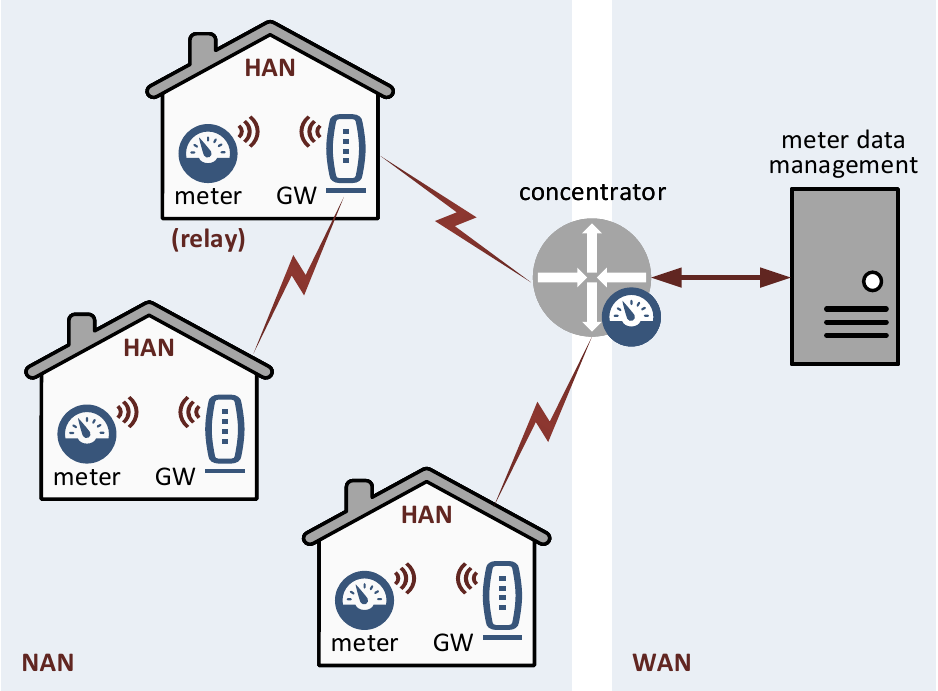}
\caption{High-level diagram of AMI architecture and scope of the security analysis}
\label{fig:ami}
\end{figure}

In this section, the architecture, components, and communication aspects of advanced metering infrastructures are provided. As illustrated in Fig. \ref{fig:ami}, an advanced metering infrastructure is comprised of three main components, namely the smart meters, the data concentrators, and the Meter Data Management (MDM) systems \cite{Hansen17}. It aims at supporting two-way communication between the various service delivery endpoints and a utility provider to allow the real-time sharing of vast amounts of data, such as power usage, outage detection, and voltage measurements, and also the remote monitoring and management (connection, disconnection, and configuration) of various services and AMI components.

\subsection{AMI components}
\label{sec:ami.components}

The AMI components are interconnected to a complex ecosystem of heterogeneous systems, covering a large geographical area, that may involve millions of smart meters and thousands of concentrators serviced by the MDM system. Details about the components are given below.

\begin{description}
\item[Smart meters.] They are installed at the service delivery points, e.g. smart homes for residencial users to measure power usage (as shown in Fig. \ref{fig:ami}), or other locations depending on the type of service a utility company is offering. A smart meter typically contains an RF component to communicate consumption data to one or more concentrators; such data may involve measurements from other IoT devices in the Home Area Network (HAN) that are obtained either directly or via the smart home's gateway \cite{k10}. The transmission of data occurs at regular time intervals called {\em duty cycles} and is less than 1KB.

\item[Concentrators.] These are the components bridging the smart meters with the MDM system, and are typically installed at electrical substations. The measurements from the smart meters in the Neighbourhood Area Network (NAN) are being collected using various topologies \cite{Hansen17,k9}, such as a point-to-multipoint or mesh topology. In the former, each smart meter communicates directly with a (single only) data concentrator, whereas in the latter, smart meters can communicate both with data concentrators and other smart meters (referred to as relays in Fig. \ref{fig:ami}) towards delivering their measurements in an efficient way. Mesh topologies are common in rural areas, where concentrators' signal range can cover a limited number of smart meters.

\item[MDM systems.] They are comprised of several subsystems to support advanced AMI operations and functionalities, including power grid management, utility optimization by means of data analytics, customer interaction and billing, etc. The distribution substations, where concentrators reside, are conneted to the utility canters by means of the public network referred to as Wide Area Network (WAN) \cite{k10,k9}.
\end{description}
As already seen above, the communication infrastructure of AMIs is divided into the HAN, NAN, and WAN layers, each one using several communications technologies to share data, as well as, to transmit and receive commands from the infrastructure's components. An overview of such technologies is provided in Table \ref{tab:comm}. The fact that power line communications require expensive equipment, and that cellular technology induces security risks due to relying on third parties, makes RF technologies to be the ideal candidates.

\begin{table}[t]
\renewcommand{\arraystretch}{1.2}
\centering
\caption{Communication technologies used in AMIs \cite{Ekanayake12}}
\label{tab:comm}
\begin{tabular}{p{.15\linewidth}p{.76\linewidth}}
\hline
{\bf Network} & {\bf Protocols} \\
\hline
HAN & Ethernet, Wi-Fi (IEEE 802.11x), Zigbee, Power line carrier (PLC), Broadband over power line (BPL) \\

NAN & Ethernet, Digital subscriber line (DSL), Frame relay, EDGE, High speed packet access (HSPA), Universal mobile telecommunications system (UMTS), Long term evolution (LTE), WiMax, 3G/4G, PLC, BPL \\

WAN & Frame relay, LTE, Multi-protocol label switching (MPLS), WiMax \\
\hline
\end{tabular}
\end{table}

\subsection{AMI requirements}
\label{sec:ami.req}

The AMIs constitute critical information infrastructures with their operation having a high impact on end-users' everyday lives \cite{Mathas20}, making their security to be of outmost importance. High-level security requirements having been reported include \cite{NIST7628,Tonyali18}:
\begin{itemize}
\item {\em Availability.} Ensure timely and reliable access to AMI data and services/power delivery.

\item {\em Integrity.} Assure that AMI data, as well as their source, have not been tampered with.

\item {\em Confidentiality.} Allowing access to AMI information only to authorized relevant entities.
\end{itemize}
As noted in \cite{NIST7628}, confidentiality is becoming more important due to increasing privacy concerns, something that led \cite{Mathas20} into defining privacy as a concrete AMI requirement to also account for inference attacks (amongst other privacy--targeting attacks). 

% Section % ================================================= |
\section{Security risks and their impact}
\label{sec:attacks}

AMIs security is considered to be one of the greatest challenges toward being accepted worldwide \cite{k1,Hansen17,k3}. Communications between AMI components incorporate real-time exchange of private and sensitive information that may include financial information of the customers \cite{k1}, vital control and safety commands and utility provider’s private information \cite{k1,Hansen17}. Moreover, AMIs are usually composed of a large number of smart meters that are generally installed in physically insecure locations and makes use of insecure wireless communication that can be easily corrupted \cite{k4}. All this makes AMIs the target of a wide variety of cyber-attacks coming from different malicious actors including illegal customers, insider attacks, criminal organisations with a large number of skilled employees, organised terrorist groups, business competitors, or even nation-states \cite{Hansen17,k4}. The system impacts of those attack range from an unforeseen peak in usage to widespread outages \cite{k1}, and in the worst scenario, a computer malware can traverse the AMI and results in millions of points of failure in a large metropolitan area, which may need several months to be fixed \cite{Hansen17,k3}. Attacks that compromise the integrity of the AMI, also have the potential to cut power from consumers, which include homeowners and other critical infrastructure such as water, hospitals and telecommunications.

The most common attacks on AMIs compromise the attack vectors that can target the AMIs end systems and communication networks \cite{Hansen17,k9}. The main goal of those attacks is to get illegal access to the devices and networks, and therefore impacts the main security goals of confidentiality, integrity and availability \cite{Hansen17,k7,k9}. This can be achieved by either physical or cyber access to the internal of the devices (e.g. smart meters, data collectors), or via a compromised supply chain \cite{Hansen17,k9}. The main risks to the AMIs security include:

\subsection{Energy theft}

Security reports consider energy theft, referred also as theft of service, as one of the most important security threats against AMIs \cite{Hansen17,Fernandes16}. This kind of attacks may be performed with a variety of known techniques. For instance, at the level of customer homes, fraudulent consumers may physically tamper with their smart meters to report malicious consumption readings, so that they are not billed for the energy they consume \cite{k3,k6}. This could be achieved by disconnecting meters from their sockets, or applying magnets to interfere with instruments, or modify the transformation ratio of the meter. It can also be performed by Cyber-attacks, which often require less expertise to execute within smart meters or via a communication link with the utility provider company (e.g. Descrambler boxes) \cite{k3}. For instance, cyber-attacks may disable the metering-related functionalities by preventing smart meters from acting on commands such as firmware updates and usage queries. This could be done by using a Denial‐of‐Service (DoS) attack on smart meter command execution \cite{k3}. 

International agencies confirm that the financial losses due to energy theft are billions of dollars per year. In this context, a world bank report found that energy theft attacks cost the industry over \$96 billion annual losses globally, with more than \$6 billion every year in the United States alone \cite{k7}.

\subsection{Data theft}

Cyber-attacks against AMIs also include illegal monitoring of sensitive data either in transit or at the AMI endpoint systems (e.g. smart meters, Data collectors, Meter Data Management System), which expose both households and utility providers to significant data theft, misinformation, and vandalism \cite{Hansen17,Fernandes16}. For instance, cyber-criminals may analysis the stolen data to reveal the electricity usage patterns and even determine the presence/absence of residents, which can strongly affect the customers' privacy and therfore thier view of deploying AMIs.

Data theft also includes unauthorized injection of data or modification of legitimate data, like unauthorized access to infrastructure configuration information and device firmware (e.g. smart meter) that can be reverse-engineered and analysed to develop attacks \cite{k7}. Data theft could also be performed by the physical theft of meters for subsequent access to the stored data \cite{k8}. Therefore, AMI requires a solid protection against data theft and unauthorized accesses by using robust security mechanisms and intrusion detection and prevention techniques.

\subsection{AMI network security concerns}
In AMIs, communication network plays a critical role in exchanging critical information between smart meters, data collectors and the utility provider such as energy consumption, pricing information, firmware updates, remote disconnects, fault or outage detection, exception messages and other parameters \cite{k10}. The last report by the Electric Power Research Institute\footnote{\url{https://www.epri.com/}} affirmed that security is one of the biggest challenges for the two-way communication path that controls the AMI network. the report stated that physically unprotected entry points and wireless networks that can be easily compromised add another attack surface to the AMI network. Therefore, the compromise of even a single vulnerable smart meter through focused attacks or reverse engineering potentially provides access to the whole AMI network. For instance, Compromised devices can be used by cybercriminals to conduct a localised denial of power by turning off power to a customer, or a group of customers, or even critical infrastructures like hospitals and telecommunications \cite{Hansen17,k10}. This can be done by sending disconnect commands to smart matters. This can also lead to a widespread denial of power when a large number, possibly millions of smart meters are disconnected \cite{Hansen17}. Such a scenario may also occur, by injecting a computer worm in the MDM system or a data collector which then propagate in the AMI network to infect other components \cite{k9,k10}. 

Another way to perform a widespread denial of power can be accomplished by using permanent denial-of-service attacks (PDoS), also known as phlashing, which can damages the smart meters so badly that it requires replacement or reinstallation of hardware. The impact of this kind of attack may need several months to replace the damaged smart meters in a metropolitan area \cite{k10}. Data injection attack is another dangerous attack that can occur in the AMI network, especially NAN, in which a compromised device tries to exhaust the bandwidth as well as the resources of its next hops \cite{k9,k10}. This can lead to data theft, loss of data integrity, denial of service, as well as full AMI system compromise. This attack could be accomplished by using compromised smart meters that legitimately participate in the routing but try to corrupt the routing function, or interfering in the routing protocols, in the NAN, by impersonating the meters \cite{k10}. In \cite{Fernandes16}, authors have been demonstrated the possibility of a command injection attack on an existing WebService SmartApp using an OAuth access token stolen from the SmartApp’s third-party Android counterpart. The signal jamming attack is one of the most basic attacks that can be made against AMI communications \cite{k8,k10}, where a signal is injected along the line path in order to prevent communication on the line. This can be done through the injection of Gaussian noise, which requires very little knowledge about the frequency at which the signal to be interfered with operates \cite{k8}. signal jamming attacks are considered as DoS attacks that would be easy to perform and difficult to detect without the appropriate countermeasures. Moreover, wireless channels used in AMI communication network, constitute a prominent target for main-in-the-middle (MitM) attacks and spoofing attacks. Cyber attacks at the smart home network level (i.e. HAN) also involves infecting connected devices by malware. In this context, several types of malware can be easily used to infect these devices and use them to spread the infection through the AMI network in the form of worms, botnets, or viruses.

\subsection{Advanced Persistent Threats (APTs)}
Modern cyber-attacks against AMI are increasingly conducted by Advanced Persistent Threats (APTs), which are very sophisticated network attacks, where experienced cybercriminals gain unauthorised access to the AMI network by using zer-day malware and stay there undetected for a long period of time \cite{k14}. APTs are usually performed by skilled groups of hackers which often utilize stealth techniques in order to remain concealed for long periods and seem to be increasing the complexity, versatility and potential damage of their attacks. Because of the high level of effort needed to perform such an attack, APTs are usually focusing on high-value targets, such as nation-states, critical infrastructures and large corporations, with the ultimate goal of stealing information over a long period of time. The Ukraine power grid attacks is an example of highly successful APTs \cite{k15}. In this incident, the cyber attack on the Ukraine's electric grid gained access to energy distribution company systems more than six months before causing the outage that temporarily left about 225,000 customers without power \cite{k15}.

GhostNet, Stuxnet, Deep Panda, APT28, APT34 and APT37 are more examples of the most destructive APTs attacks in last years. The main issue of APT attacks is that even when they are discovered and the immediate threat appears to be covered, the cybercriminals may have left many backdoors open that enable them to return when they want. Additionally, traditional cyber defences, such as antivirus and signature-based intrusion detection systems, are unable to protect against these types of attacks. Therefore, monitoring of traffic, users and entities behaviour can greatly help identify penetrations, lateral movement and exfiltration at different stages of an APT attack, which is the purpose of this work.

% Section % ================================================= |
\section{ML-based malware detection}
\label{sec:malware}
\begin{figure}[t]
\centering
\includegraphics[width=0.75\linewidth]{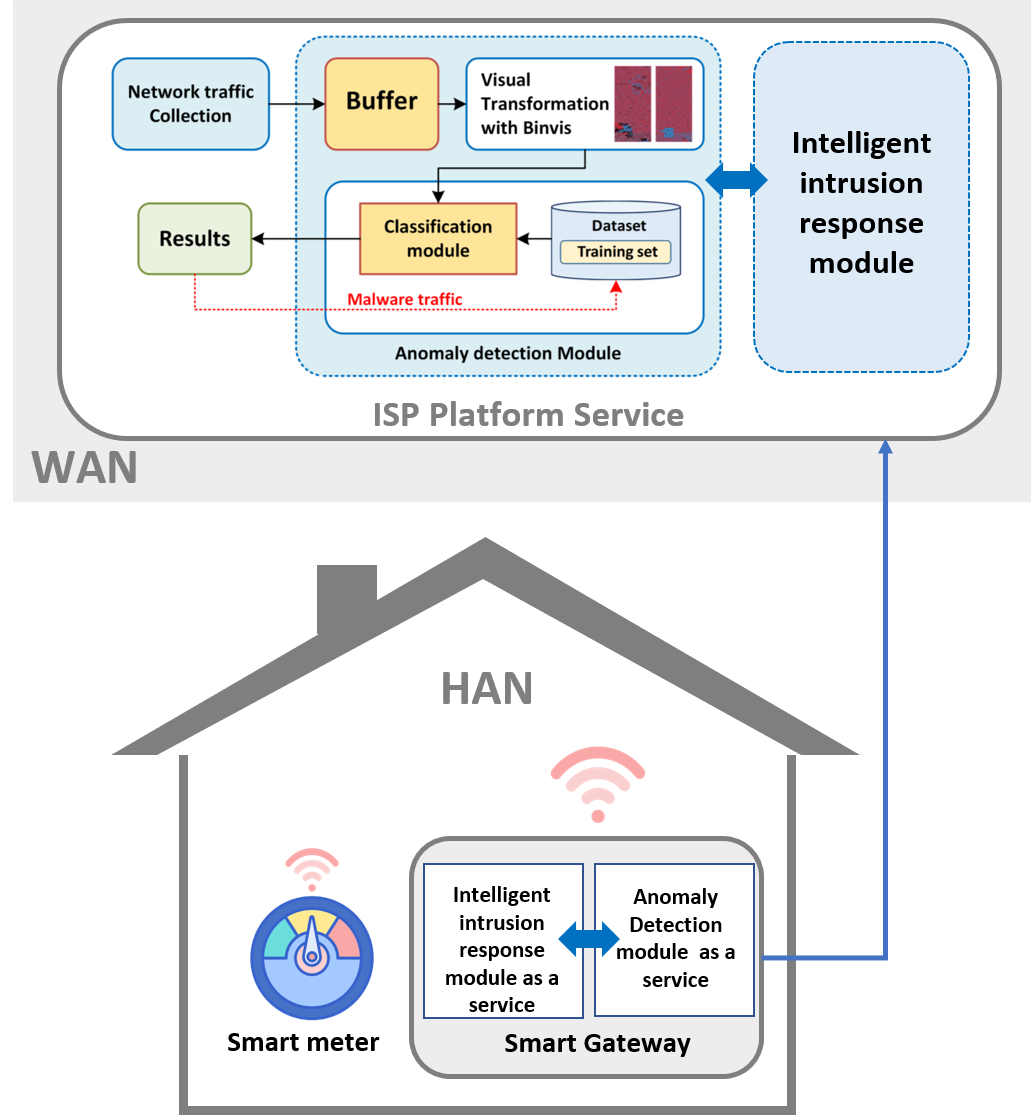}
\caption{High-level architecture of the proposed approach}
\label{fig:proposed.approach}
\end{figure}

This section presents the methodology used for developing the proposed ML-based malware detection system. The main objective of this system is to defend the whole metering infrastructure from malicious attacks using a novel intrusion prevention technique based on machine learning and binary visualisation. The proposed approach converts incoming network traffic into RGB images by using the visual representation tool Binvis\footnote{\url{https://binvis.io/}}. Then, the produced images are analysed and classified using a learning algorithm. different learning algorithms can be used for classifying the produced images like Residual Neural Network (ResNet50), Mobilenet, Self-Organizing Incremental Neural Networks (SOINN). 
The network traffic collection is done by using pcap files containing pre-captured network traffic (i.e. normal and abnormal traffic) that can be replayed to collect it again. Then, received data is stored out to a file that contains the data from the payload in the packet, so the visual representation tool can plot it into a 2D image. As illustrated in Figure \ref{fig:proposed.approach}, the detection and mitigation of potential cyber-attacks is performed at the networks level as well as at the device levels. The device monitoring is done at the gateway which is running the proposed ML-detection approach as a service due to the limited processing resource at the smart home gateway. Similarly, at the network level, the ML-detection approach, which is integrated to the intrusion detection system, is used to monitor the incoming and outgoing network traffic. This helps to enforce network mitigation by applying the mitigation and remediation actions as necessary when an attack is detected. 

\subsection{Binary visualisation}
The research community has started considering the concept of image visualization for malware analysis and detection, which can successfully handle obfuscated and zero-day malware \cite{k12,k13}. This technique has proven to be effective because it leverages the structural similarity between known and new malware binaries. Moreover, visual analysis helps analysts to accurately capture and highlight malicious behaviour of malware samples, thus helping increase the efficiency of the detection system \cite{k14}. Most of these techniques transform malware detection into an image classification problem so that can be processed by machine learning algorithms \cite{k12,k13,k14}.

\begin{figure}[t]
\centering
\includegraphics[width=0.75\linewidth]{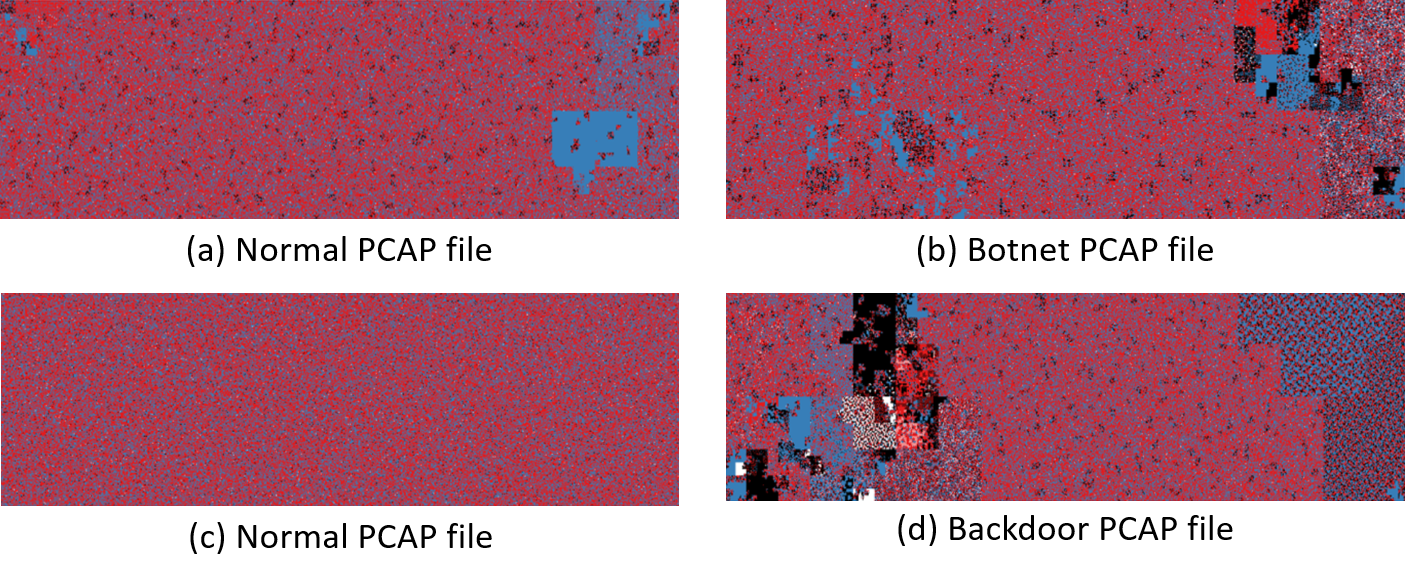}
\caption{Binvis images of normal and malware pcap files created with the Hilbert space-filling curve}
\label{fig:pcap}
\end{figure}

In our approach, the binary content of the input file are seen as a byte string, where each byte’s value is mapped to a colour based on the equivalent value in the ASCII table. Binvis divided the different ASCII bytes into four groups of colours, where red colour is attributed to extended ASCII bytes, blue colour is assigned to Printable ASCII bytes and green colour is assigned to control bytes. Black (0x00) and white (0xFF) colour respectively represent null and (non-breaking) spaces. Then, the coordinates of each byte colour in the output image are identified by using the clustering algorithm’s space-filling curves (Figure. \ref{fig:pcap}). The size of the output RGB image is 784 ($1024\times 256$) bytes. 

\subsection{Malware detection}
\label{sec:ML}
In the context of malware detection and analysis, Machine Learning has recently gained significant attention for its capability to accurately detect malware attacks and therefore reduce the false positive alarms by proactively reacting against unknown attacks. Supervised learning algorithms can be used to analyse the available information of the system activity (e.g. network trafﬁc), by using different features derived from dynamic analysis of the malware. Then, use extracted features to train the learning model to detect potential attacks. The output results are usually presented in a binary form (i.e. normal or malware), and each data sample is labelled as either normal or anomaly \cite{k35,k36,k41}. In this context, the predictive accuracy of various supervised learning algorithms has been tested like the Naive Bayes (NB), K-nearest Neighbour (KNN), Decision Tree (J48), Multi-Layer Perceptron (MLP) and Random Forest (RF) and Support Vector Machine (SVM). Experimental results noted that most of the learning algorithms gave a satisfying accuracy of over 90\%, with low rates of false positives \cite{k38}.

On the other hand, unsupervised learning algorithms, learn what is considered as normal, and then apply statistical tests to determine if a specific activity is an anomaly. A system based on this kind of anomaly detection method could detect any type of anomaly, including unknown and new attacks \cite{k35,k42}. In the last few years, several unsupervised learning algorithms, especially Deep Learning techniques, which represent a huge step forward for unsupervised learning, have been employed for intrusion detection \cite{k38}. Such as Restricted Boltzmann Machine (RBM), Self-Organizing Incremental Neural Networks (SOINN) \cite{k41,k42}, deep belief network (DBN), Residual Neural Network (ResNet), Deep Neural Network (DNN), Recurrent Neural Network (RNN), etc. Most of these techniques transform malware detection into an image classification problem so that can be processed by the learning algorithms. For instance, the STAMINA (Static Malware-as-Image Network Analysis) malware detection approach \cite{k40}, which is recently proposed by Microsft and Intel, converts input binary files into grayscale images then, a trained neural network classifier is used to analysis and classify the output images as legitimate or malware. The learning algorithm is trained on a huge amount of real-world data (2.2 million PE (Portable Executable) file hashes) that Microsoft has collected from Windows Defenders installations. STAMINA has proven to be effective, with over 99.00\% accuracy in classifying malware and a false positive rate slightly under 2.6\%. However, this approach works well with small files, but it struggles with larger ones.        

In our approach, the produced Binvis images will be analysed using a trained learning algorithm to perform the classification of the incoming traffic as normal or malware (see Fig. \ref{fig:proposed.approach}). Detected malware traffic will be used to continuously train the classifier in order to enhance their detection accuracy. For that, the learning algorithm will be trained on a dataset that was created in the Cyber-Trust project testbed. The dataset includes a mixture of 2D images of normal and malware trafﬁc that were collected from different network trafﬁc sources. Normal PCAP ﬁles contain normal trafﬁc captured from various clean devices in the Cyber-Trust project network and other sources. While malicious pcap ﬁles were collected from different public sources of malware PCAP ﬁles including the malware trafﬁc analysis repository\footnote{\url{https://github.com/tatsui-geek/malware-traffic-analysis.net}}, the NETRESEC repository\footnote{\url{https://www.netresec.com/?page=PcapFiles}} and the malware datasets of the stratosphere lab\footnote{\url{https://www.stratosphereips.org/datasets-malware}}. The malware pcap ﬁles contain real malicious trafﬁc that was generated by different types of attacks such as trojans, botnets, IoT based attacks (DDoS, Key loggers, OS scans, spyware), backdoors, etc.

% Section % ================================================= |
\section{Intelligent intrusion response}
\label{sec:irs}

Cyber-attacks against AMIs constitute a major threat and thus much research has been devoted to their study with the upshot of developing an effective Intrusion Response System (IRS). In turn, this requires accurately modeling the cyber-attacks themselves, the potential attackers' behaviors, and the available defensive strategies. Hereinafter, we unveil the basic methodologies that are utilized towards the design of {\em Intelligent IRS} capable of optimally mitigating AMI cyber-attacks.

\subsection{Graphical network security models}
\label{sec:gnsm}

The use of a Graphical Network Security Model (GNSM) is among the most common methodologies adopted for analyzing network security. Many different models have been proposed, but they can be divided into tree-based and graph-based models. The main difference between them is that the former are being used to describe a single attack goal, while the latter can present scenarios with multiple attack goals. In addition, attack trees focus on the consequence of an attack, while attack graphs typically focus on the attackers’ activities and how they interact with the targeted infrastructure. Therefore, if there is a need to capture the attack paths, then a graph-based model would be preferable. On the other hand, if the focus is the assessment of the overall network security, where only the most critical vulnerabilities of the system need to be analyzed, then a tree-based model would be more suitable.

\begin{figure}[t]
\centering
\includegraphics[width=0.75\linewidth]{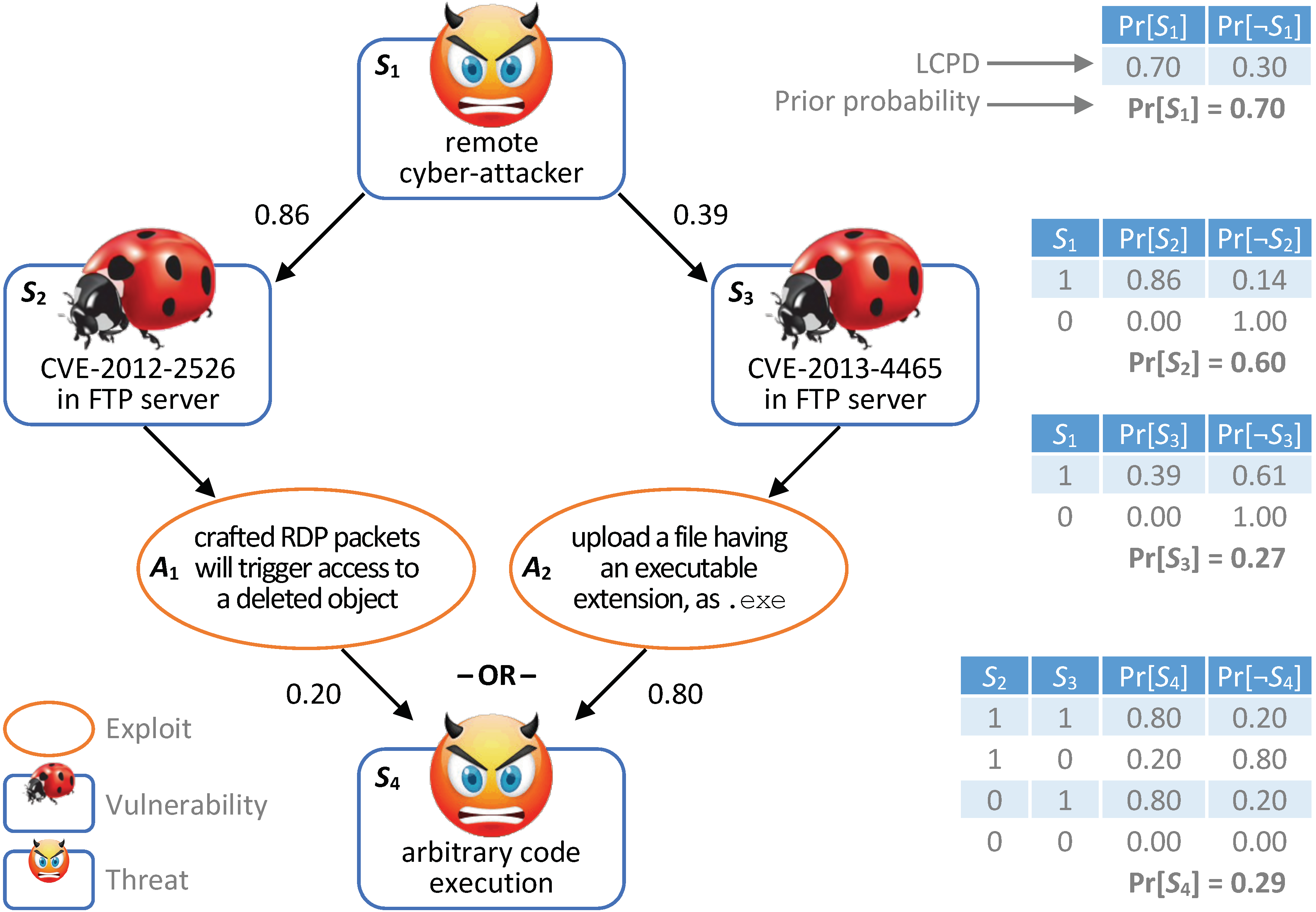}
\caption{Bayesian attack graph illustrating the computation of local conditional probability distributions (the example is adjusted from \cite{Poolsappasit12})}
\label{fig:graph}
\end{figure}

Attack Graphs (AG) have been employed in formal risk/threat analyses of large networks by a number of authors \cite{Poolsappasit12}. An AG represents the attack states and the transitions between them as shown in the example of Fig. \ref{fig:graph}. AGs can be used to identify attack paths that are most likely to succeed, or to simulate various attacks. In AGs a node represents states (e.g. host, privilege, exploit or vulnerability), and an edge is a directed transition from a pre-condition to a post-condition when an event of the state has been executed. A Bayesian attack graph (BAG), as the example illustrated in Fig. \ref{fig:graph}, is an important instance of AGs. A BAG can be seen as a {\em directed acyclic graph} over vertices representing random variables and edges signifying conditional dependencies between pairs of vertices; thus it is very convenient for conducting a probabilistic analysis of the attacks.

\subsection{Attack graph generation}
\label{sec:irg}
Various ways to model network topology information and generate an AG have been proposed in the literature. For large networks, like those corresponding to AMIs, a non-automated AG generation is impossible as the resulting graph would have a vast number of vertices. On the other hand, an automated AG generation process should be both {\em exhaustive} (all possible attacks are modelled) and {\em succinct} (only the network states from which an attacker can reach his goal are contained) so as to be efficient. The following aspects concerning the AG generation process are relevant.
\begin{enumerate}
\item {\em Reachability analysis:} how AMI network interconnectivity is modeled across all layers of the open systems interconnection (OSI) model, and how the calculation of the possible ways an attacker can reach the goal state is performed.
\item {\em Template determination:} how the relations between the required privileges to exploit a vulnerability (pre-conditions) and the privileges gained after a successful vulnerability exploitation (post-conditions) are being modeled.
\item {\em Structure determination:} how the actual representation of the AG is defined and how expressive is the information collected (e.g. to subsequently allow performing risk analysis, computing the optimal remediation action, etc.).
\item {\em Core building mechanism:} how the algorithms are employed to build the AG, i.e. to discover all attack paths from the initial states an attacker may start to the chosen target states.
\end{enumerate}
For the development of the {\em Intelligent IRS} AG Generator (iRG) of Fig. \ref{fig:proposed.approach}, the Multi-host, Multi-stage Vulnerability Analysis Language (MulVAL) was used as a reasoning system in order to model AMI networks and generate a type of AGs, referred to as Logical Attack Graphs (LAG). These were subsequently mapped into BAGs to perform probabilistic modelling of the attacks.

Initially, output from the supported vulnerability scanning tools (e.g. OpenVAS and Nessus), as well as network topology information, are expressed as Datalog tuples, which are then processed by the reasoning engine. To combat issues related to the poor scalability of AG generation (most approaches have exponential complexity), the {\em monotonicity assumption} on the attacker’s behavior has been made; more precisely, we have assumed that the attacker will not give up any previously attained capabilities. Under this assumption, the AG generation reduces to polynomial complexity.

\begin{figure}[t]
\centering
\includegraphics[width=0.75\linewidth]{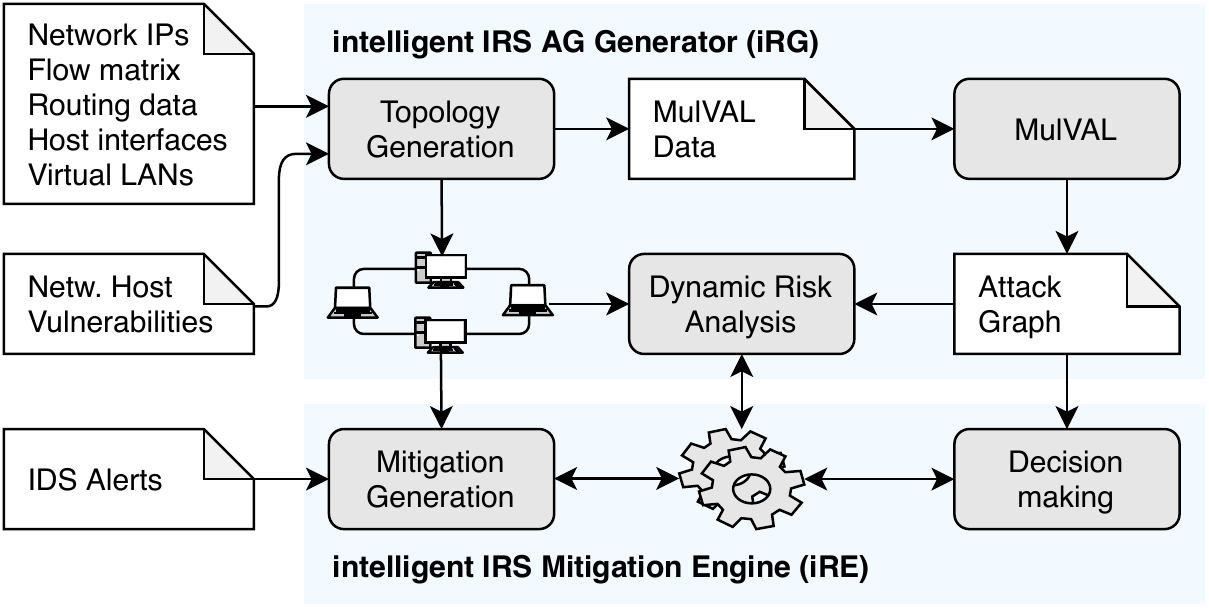}
\caption{High-level architecture of the iIRS}
\label{fig:iirs}
\end{figure}

\subsection{Decision-making engine and mitigation}
\label{sec:ire}

In this section, we go a step further and deal with the intelligent intrusion response process, where the defender has to decide how to react against an attacker. The theoretical approach taken by this work relies on Game Theory (GT) that further leverages the representation offered by the GNSMs mentioned above. In our model, the attacker aims at exploiting system vulnerabilities for progressing his attack in an AMI with the aim of reaching some goal state, while the defender aims at simultaneously preventing the attacker's progression and maintaining the AMI's availability and other security requirements. Our goal is to develop an IRS that is capable of optimally responding to intrusions; this is referred to as the {\em Intelligent IRS} Mitigation Engine (iRE).

The high-level architecture of the Intelligent IRS, with the components having been mentioned, is illustrated in Fig. \ref{fig:iirs}. This also depicts the Decision-Making part of the iRE that relies on GT to yield the optimal defensive actions (subsequently being translated into applied mitigation rules). As in \cite{Miehling18}, we also employ a Partially Observable Monte-Carlo Planning (POMCP) algorithm to simulate possible future state trajectories from the current belief state (i.e. the defender's view of the AMI network's security) in order to evaluate the effectiveness of various defense decisions made, thus enabling the defender to make a selection in real-time.

The intrusions are being signaled by the alerts generated from the IDS of Fig. \ref{fig:proposed.approach}, while the responses take the form of IDS and firewall rules. An efficient algorithm was implemented for generating the mitigation actions and for temporarily changing the AG by modifying the AMI's attack surface. This is achieved by changing the connectivity of network hosts, thus effectively blocking access to vulnerable services and/or devices by employing the generated rules. The algorithm starts with a desired node to be blocked (corresponding to a security condition) and moves towards the leaves of the AG. It explores (using depth-first search) information that might be available to AG vertices for generating firewall rules and stores the connections and relations between rules in a tree structure. This structure represents multiple sets of firewall rules to be applied in order to block the progression of an attacker towards a goal condition of the AMI network's AG. In principle, the goal of an attacker is linked with the desired ability to execute arbitrary code at a specific AMI device. This is defined as follows
\vspace*{2pt}
\begin{verbatim}
   execCode(_attacker, _host, _permission)
   execCode(_host, _permission)
\end{verbatim}
\vspace*{2pt}
where arguments beginning with an underscore represent variables. This allows the iRE's Decision-Making to weight differently cases where an attacker is believed to be close to a target condition, thus yielding excellent intrusion mitigation performance.

% Section % ================================================= |
\section{Conclusions}
\label{sec:conclusions}
As can be concluded from this paper, the combination of ML Intrusion Detection Systems (IDS) and Graphical Cyber Security Models(GCSMs) can lead to an innovative class of intelligent intrusion response systems (iIRS) providing dynamic security risk assessment and intelligent mitigation strategies to defend against adaptive multi–stage cyber-attacks on AMI, in an optimal and autonomous fashion\cite{Miehling}. This can be achieved by building upon advanced game-theoretic security approaches, where accurate model of attackers and defenders (players), their interactions and  the AMI network parameters would be able to calculate all the possible scenarios and provide the optimal solution to be applied by IDS. This will generate a positive impact on AMIs, small and medium-sized enterprises, but also to critical infrastructures and industrial IoT facilities as will be able to mitigate even (unknown) sophisticated cyber-attacks.

\section{acknowledgement}

\begin{wrapfigure}{L}{0.1\linewidth}\centering
\vspace{-14pt}\hspace{-5pt}
\includegraphics[width=.58in]{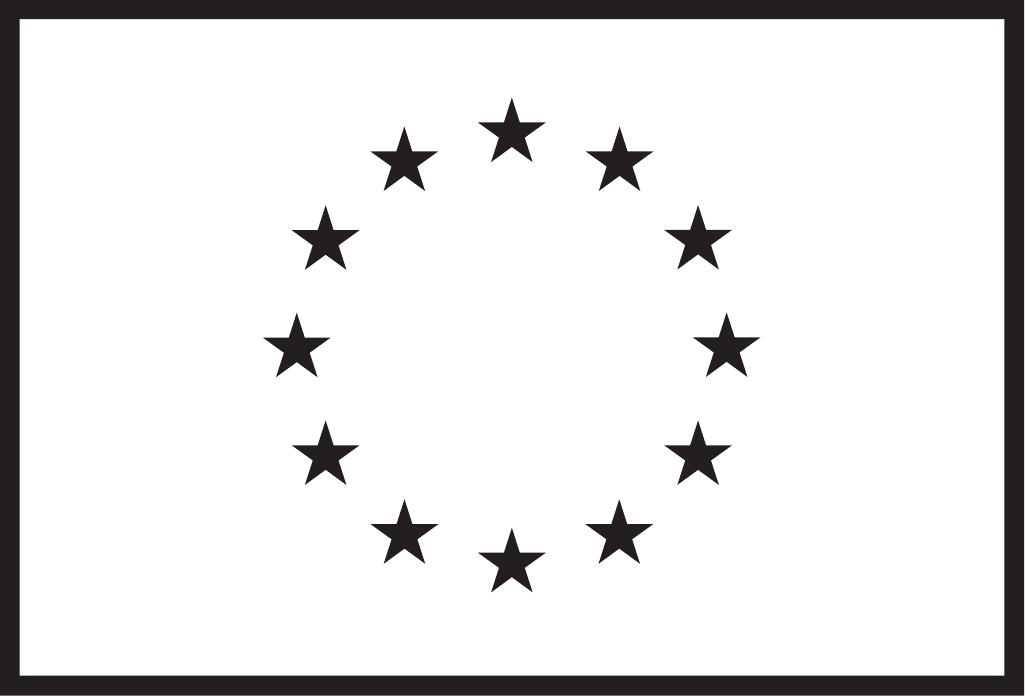}
\vspace{-24pt}
\end{wrapfigure}
This project has received funding from the European Union's Horizon 2020 research and innovation programme under grant agreement no. 786698 and 833673. The work reflects only the authors' view and the Agency is not responsible for any use that may be made of the information it contains.

%%
%% The next two lines define the bibliography style to be used, and the bibliography file.

\end{document}